%% file: main.tex
\documentclass[10pt]{article}
\usepackage{preprint} 
\usepackage{microtype}
\usepackage{amsmath,amssymb}
\usepackage{algorithm}
\usepackage{algpseudocode}
\usepackage{multirow}
\usepackage{float}
\usepackage{tikz}
\usetikzlibrary{arrows.meta}
\usepackage{hyperref}
\usepackage[capitalise,noabbrev,nameinlink]{cleveref}

\logos{%
  \mbox{%
    \raisebox{-0.5\height}{%
      \includegraphics[height=0.30in,
        trim=465bp 80bp 490bp 75bp,clip]{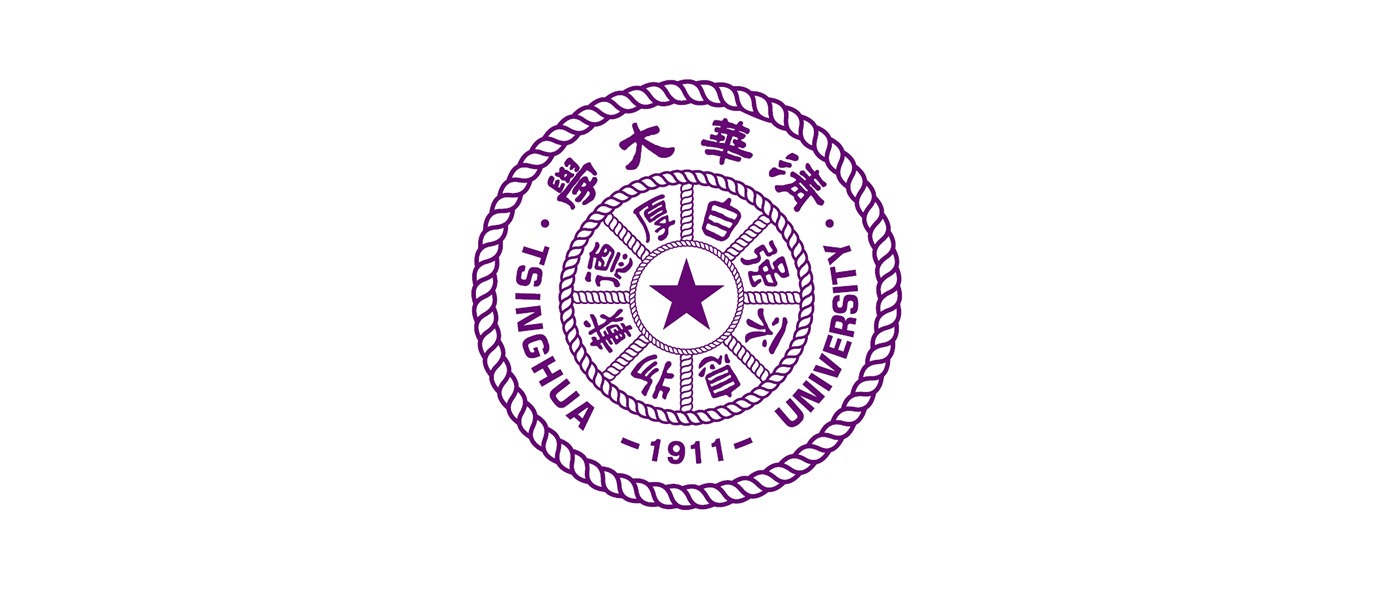}}%
    \hspace{7pt}%
    \raisebox{-0.5\height}{%
      \includegraphics[height=0.32in,
        trim=0bp 0bp 0bp 120bp,clip]{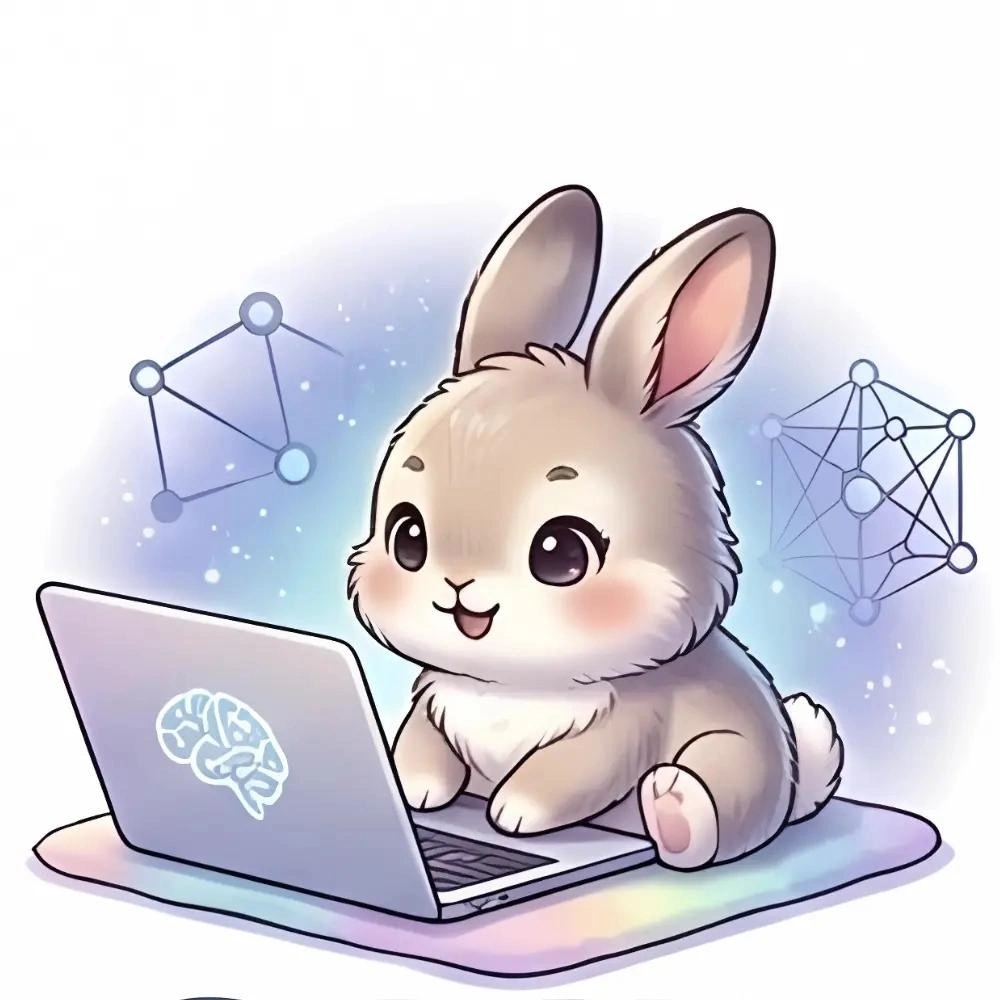}}%
    \hspace{7pt}%
    \raisebox{-0.5\height}{%
      \fontfamily{Montserrat-LF}\fontseries{medium}\fontshape{n}%
      \fontsize{10.5}{12}\selectfont\color{accent}%
      \textls[10]{\groupname}}%
  }%
}

\title{Can Coding Agents Sustain Continual Evolution without Forgetting?}
\author[1]{Shangziqi Zhao}
\author[1]{Yitong Zhang}
\author[1,\dagger]{Jia Li}
\affiliation[1]{Tsinghua University}
\date{\today}
\metadata[First author]{\email{szqzhao618@gmail.com}}
\metadata[Corresponding author]{$^{\dagger}$\,\email{jia\_li@mail.tsinghua.edu.cn}}

\begin{document}
\maketitle

\input{sections/0_abstract}
\input{sections/1_introduction}

\input{sections/2_problem_definition}
\input{sections/3_experimental_setup}
\input{sections/4_experimantal_results}

\input{sections/5_related_work}
\input{sections/6_conclusion}

\bibliographystyle{preprint}
\bibliography{main}

\clearpage
\appendix
\input{appendix/1_implementation_details}

\input{appendix/2_discussions}

\end{document}

%% file: sections/0_abstract.tex
\begin{abstract}
Recently, harness self-evolution has emerged as an important paradigm for improving agent performance. Prior studies show that evolving the agent harness can effectively enhance agent capabilities without modifying the foundation model's parameters. We find that existing approaches largely study agent evolution on a single task. In practice, coding agents may need to handle multiple task types over time. Whether an agent harness can continuously evolve across multiple tasks without suffering from catastrophic forgetting remains an important open question.

In this paper, we investigate this open question through a comprehensive empirical study of continual harness self-evolution. We organize three software engineering task datasets into six different evolution directions and apply two representative harness evolution methods to perform continual evolution along these directions. Our experimental results show that evolution on a subsequent dataset can substantially undermine the performance on previously seen datasets. For example, with DeepSeek-V4-Flash-0731, Meta-Harness improves its performance on SWE-Bench to 94.0\% after evolution. However, after further evolution on FeatureBench, its performance on SWE-Bench drops to 32.0\% by the fifth round, corresponding to a 66\% relative decrease. Finally, based on our experimental findings, we identify several insights, providing directions for the continual self-evolution of agent harnesses. Code is available at \url{https://github.com/THU-Agent/continual-harness-evolution}.
\end{abstract}

%% file: sections/1_introduction.tex
\section{Introduction}
\label{sec:introduction}

\begin{figure*}[t]
  \centering
  \includegraphics[width=\linewidth]{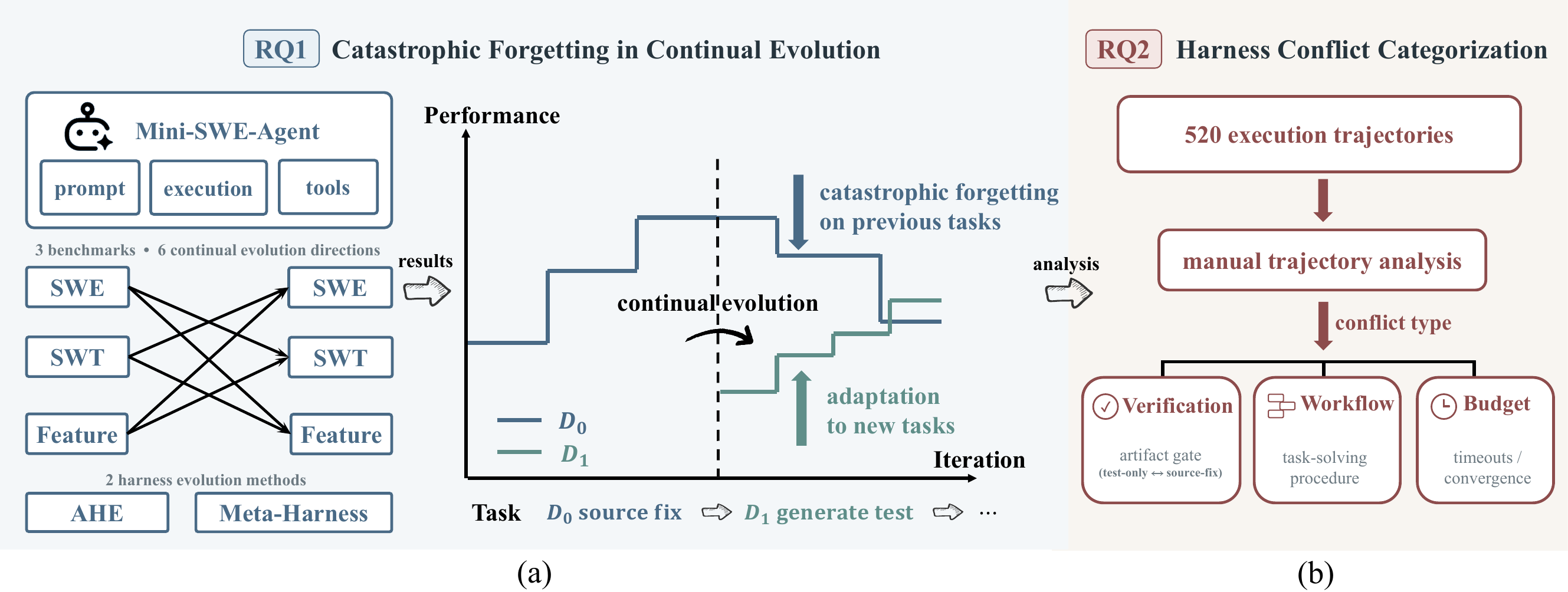}
  \caption{Overview of our study on continual harness self-evolution.}
  \label{fig:intro}
\end{figure*}

Coding agents have been increasingly used in software engineering tasks~\citep{DBLP:journals/corr/abs-2409-02977}, such as issue solving, feature addition, and test generation~\citep{DBLP:conf/nips/YangJWLYNP24, DBLP:conf/iclr/0001LSXTZPSLSTL25, DBLP:conf/nips/0003ZWZ0C24, DBLP:conf/icse/BouzeniaDP25}. A coding agent typically consists of a foundation model and an agent harness~\citep{DBLP:journals/corr/abs-2605-18747, DBLP:conf/nips/YangJWLYNP24}. The harness coordinates the interaction between the model, tools, and environment, thereby influencing how the agent perceives task information, invokes tools, and executes tasks. Different harnesses can substantially affect agent performance. In our experiments, with the foundation model fixed, different harnesses achieve substantially different performance on SWE-Bench, ranging from 82.0\% to 96.0\%.

To improve the agent harness, recent studies have explored harness self-evolution, which enables agents to learn from their own execution experience by analyzing trajectories, identifying strategies, and using these insights to automatically improve their runtime systems~\citep{jiang2026selfimprovingagents}. Recently, a number of approaches have been proposed, demonstrating promising improvements in agent capabilities~\citep{DBLP:conf/iclr/AgrawalTSZKOSSR26, DBLP:conf/iclr/ZhangHUMHKRWJLT26, DBLP:journals/corr/abs-2508-16153, DBLP:conf/ecai/ChhikaraKASY25, DBLP:journals/corr/abs-2604-01687, DBLP:journals/corr/abs-2602-08234, DBLP:journals/corr/abs-2604-25850, DBLP:journals/corr/abs-2603-28052, DBLP:journals/corr/abs-2606-09498, DBLP:journals/corr/abs-2606-14249}. For example, AHE~\citep{DBLP:journals/corr/abs-2604-25850} improves the performance of the seed harness $NexAU_0$ on Terminal-Bench 2 from 69.7\% to 77.0\% after ten rounds of evolution.

However, existing harness self-evolution studies primarily focus on independently evolving harnesses for a single kind of task. In practical scenarios, coding agents are expected to address various software engineering tasks. The Stack Overflow Developer Survey 2026 report that 72.9\% of developers use AI agents for writing, editing, or generating code, 62.2\% for debugging or troubleshooting, and 50.6\% for writing or maintaining tests~\cite{stackoverflow2026agenttasks}. In software development, a user may first ask a coding agent to add a new feature and later ask it to write test files. Inspired by this observation, we propose \textit{continual harness self-evolution}, which refers to successively evolving a harness across different task types, in contrast to evolving a harness independently for a single task. However, to the best of our knowledge, no existing work has explored continual evolution of coding agent harnesses across different kinds of tasks.

To bridge the above research gap, we conduct a systematic study of continual harness self-evolution for coding agents. Our study aims to answer the following research questions:

\textbf{RQ1: Can existing harness self-evolution methods sustain continual evolution across multiple software engineering tasks?}
Continual harness self-evolution requires an agent to keep improving on each new tasks without losing the capabilities it acquired on previous ones. To examine whether existing methods meet this requirement, we construct a continual evolution setting in which the harness is evolved on a sequence of software engineering task types rather than on a single task. Specifically, we evaluate two representative harness self-evolution methods, AHE~\citep{DBLP:journals/corr/abs-2604-25850} and Meta-Harness~\citep{DBLP:journals/corr/abs-2603-28052}, across three software engineering datasets: SWE-Bench~\citep{DBLP:conf/iclr/JimenezYWYPPN24} for code repair, SWT-Bench~\citep{DBLP:conf/nips/MundlerMHV24} for test case generation, and FeatureBench~\citep{DBLP:conf/iclr/ZhouZWHWHYWPFTZ26} for feature generation. Each continual evolution experiment forms a two-stage task sequence, where the harness first evolves on an earlier dataset and then continues evolving on a later one, allowing us to track how performance on the earlier task changes as evolution proceeds. The results reveal substantial catastrophic forgetting during continual harness self-evolution, as shown in Figure~\ref{fig:intro} (a). For example, Meta-Harness improves SWT-Bench to 100.0\%, but its performance on SWT-Bench drops to 62.0\% after further evolution on FeatureBench.
 
\textbf{RQ2: What are the root causes of catastrophic forgetting in the harness?}
As shown in Figure~\ref{fig:intro} (b), to understand why continual harness self-evolution can lead to catastrophic forgetting, we collect over 520 agent trajectories and manually analyze the harness modifications introduced during continual harness self-evolution. We categorize these modifications into three major types of harness-level conflicts: verification, workflow, and budget strategy conflicts. Our ablation experiments show that different conflict types have substantially different effects on agent performance. We hope these findings shed light on the design of future self-evolution methods that are aware of such conflicts, and we call for more attention to this overlooked yet critical issue. \\

The main contributions of our work are as follows:
\begin{contributions}
    \item Motivated by the practical use of coding agents across diverse tasks, we propose the problem of harness continual self-evolution and identify catastrophic forgetting as a critical challenge.
    \item We evaluate existing harness self-evolution methods under continual evolution, revealing catastrophic forgetting on previous acquired capabilities.
    \item We manually analyze agent trajectories and identify key harness-level conflicts that contribute to forgetting phenomena.
\end{contributions}

%% file: sections/2_problem_definition.tex
\section{Problem Definition}
\label{sec:problem_definition}

A coding agent consists of a foundation model and an agent harness that coordinates its interaction with the software environment~\citep{DBLP:journals/corr/abs-2605-18747, DBLP:conf/nips/YangJWLYNP24}. Recent work has explored \textit{harness self-evolution}, which improves agent capabilities by modifying the harness based on execution experience while keeping the foundation model parameters fixed~\citep{jiang2026selfimprovingagents}. Formally, an agent is represented as $A = \langle M, H \rangle$, where $M$ denotes the foundation model and $H$ denotes the harness. 
Given execution trajectories $\mathcal{T}$ collected from a dataset $\mathcal{D}$, a self-evolution method $\Omega$ updates the harness as
$H' = \Omega(H, \mathcal{T})$.

In practical settings, coding agents may need to evolve across different software engineering task types, such as issue solving, feature addition, and test generation. We refer to this setting as \textbf{continual harness self-evolution}. Given an ordered sequence of datasets $\mathcal{D}=(D_0,D_1,\ldots,D_K)$, where each dataset represents a kind of software engineering task, the harness is updated repeatedly as follows:
\begin{equation}
H_{i+1} = \Omega\left(H_{i}, \mathcal{T}_i\right),
\end{equation}
where $\mathcal{T}_i$ denotes the execution trajectories collected from tasks in $D_i$, and $i=0,\ldots,K$. 

A central challenge in continual harness self-evolution is to adapt to the new task type without substantially degrading performance on the previously encountered one. This challenge is related to the stability--plasticity dilemma in continual learning~\citep{DBLP:journals/cvgip/CarpenterG87,DBLP:journals/nn/YangWWD26}. When adaptation to new experience compromises previously acquired capabilities, \textit{catastrophic forgetting} may occur~\citep{DBLP:journals/corr/KirkpatrickPRVD16}. In this paper, we define catastrophic forgetting as a substantial performance degradation on the source task after further harness evolution on the target task.

Ideally, continual harness self-evolution should improve performance on each new dataset while preserving performance on previously encountered datasets. Specifically, after evolving on $D_i$, the resulting harness $H_{i+1}$ should not degrade performance on dataset $D_{j}$, where $0 \leq j < i \leq K$:
\begin{equation}
\mathrm{Eval}(H_{i + 1}, D_i)
>
\mathrm{Eval}(H_i, D_i),
\end{equation}
\begin{equation}
\mathrm{Eval}(H_{i + 1}, D_{j})
\geq
\mathrm{Eval}(H_{j + 1}, D_j).
\end{equation}
Here, $\mathrm{Eval}(H,D)$ denotes the performance of an agent with harness $H$ and a fixed foundation model on dataset $D$. The first condition captures adaptation to the current task, while the second captures retention of previous-task performance.

To quantify catastrophic forgetting, we measure the absolute and relative performance drops on the previous dataset:
\begin{equation}
\Delta_{\mathrm{forget}}
=
\mathrm{Eval}(H_{j + 1}, D_j)
-
\mathrm{Eval}(H_{i + 1}, D_j),
\label{eq:delta}
\end{equation}
\begin{equation}
\rho_{\mathrm{forget}}
=
\frac{
\mathrm{Eval}(H_{j + 1}, D_j)
-
\mathrm{Eval}(H_{i + 1}, D_j)
}{
\mathrm{Eval}(H_{j + 1}, D_j)
}.
\label{eq:rho}
\end{equation}

%% file: sections/3_experimental_setup.tex
\section{Experimental Setup}
\label{sec:exp_eval}

With the problem defined in Section~\ref{sec:problem_definition}, we conduct a study of continual harness self-evolution. In this section, we introduce the datasets, coding agent framework, self-evolution methods, and implementation details used in our experiments.

\subsection{Datasets}
We study the continual harness self-evolution between different types of software engineering tasks, including code repair, test case generation, and code generation.

SWE-bench Verified~\citep{DBLP:conf/iclr/JimenezYWYPPN24} is one of the most widely used code repair benchmarks in software engineering. It contains 500 high-quality instances derived from real-world software issues collected from GitHub and manually verified by human annotators. The goal of each task is to identify and resolve the underlying issue by modifying the corresponding repository.

SWT-Bench~\citep{DBLP:conf/nips/MundlerMHV24} focuses on evaluating agents’ ability to generate test cases for real-world software issues. Each task provides the agent with source code and an issue report collected from GitHub, and the generated tests are expected to fail on the original implementation but pass after the corresponding fix is applied. Notably, all task prompts originally follow an identical sentence structure. To avoid potential shortcuts and improve consistency with other datasets, we paraphrase this structure while preserving its semantics, resulting in diverse surface forms across tasks, as discussed in Appendix ~\ref{app:disc}.

FeatureBench~\citep{DBLP:conf/iclr/ZhouZWHWHYWPFTZ26} evaluates agents on implementing new features in real-world software repositories. Given a feature requirement and its interface description, the agent is expected to implement the requested functionality and place the resulting implementation in the testbed directory. The evaluation considers both fail-to-pass (F2P) and pass-to-pass (P2P) tests: F2P tests assess whether the new feature is correctly implemented, while P2P tests verify that the modification does not break existing functionality.

Due to resource constraints, we randomly sampled a limited number of instances from each dataset for self-evolution, including 50 instances from SWE-Bench Verified, 50 from SWT-Bench, and 30 from FeatureBench.

\subsection{Initial Agent Harness}
We initialize the harness with Mini-SWE-Agent~\citep{DBLP:conf/nips/YangJWLYNP24}, a commonly used agent framework for software engineering tasks. Its core implementation consists of only around 100 lines of Python code, while achieving a score of over 74\% on SWE-bench Verified. The original implementation provides the agent control flow, model interface, Bash tool definition, and configurations for both local and Docker-based environments. To prevent the evolver from modifying the interface specifications or execution environment, we restrict the scope of self-evolution to several aspects: modifying the system prompt, modifying the agent execution logic, and adding or modifying tools.

\subsection{Evaluated evolution algorithms}
We select two representative agent harness self-evolution methods to evaluate their catastrophic forgetting during continual evolution.

AHE~\citep{DBLP:journals/corr/abs-2604-25850} constructs a closed-loop self-evolution process. The coding agent interacts with the environment through the current harness and generates trajectories on tasks from the target dataset. The Agent Debugger then accesses the trajectory files using general-purpose shell commands and scripting tools, and summarizes the root causes of successful and failed trajectories. Finally, the Evolve Agent uses the evidence produced by the Agent Debugger to modify the harness code, preparing the updated harness for the next round of evolution.

Meta-harness~\citep{DBLP:journals/corr/abs-2603-28052} continuously maintains a Pareto frontier of harnesses in contrast. In each round of evolution, the coding agent proposes new modifications by referring to existing harnesses and accumulated experience. The newly modified harnesses are then evaluated on tasks from the current dataset, with the evaluation logs stored in the file system for subsequent evolution.

Since neither method’s publicly available implementation supports the evolution of Mini-SWE-Agent, we implement the corresponding functionality based on their original codebases, while aligning our implementations with the original experimental settings as closely as possible.

\subsection{Implementation Details}
For the hyperparameters, we set the rollout temperature to 0. For AHE and Meta-Harness, we follow the settings of their original papers and perform 9 and 5 rounds of evolution, respectively. In each round, the same batch of tasks is executed twice, and the resulting trajectories are analyzed to guide the evolution.

We use DeepSeek-V4-Flash-0731~\cite{DBLP:journals/corr/abs-2606-19348} as the foundation model for both task execution and harness evolution. To ensure a consistent execution environment across all experiments, we uniformly use E2B as the sandbox environment. We set the time limit for each task execution to 1 hour. 

%% file: sections/4_experimantal_results.tex
\section{Experimental Results}
\label{sec:exp_res}

\begin{figure*}[t]
  \centering
  \includegraphics[width=\linewidth]{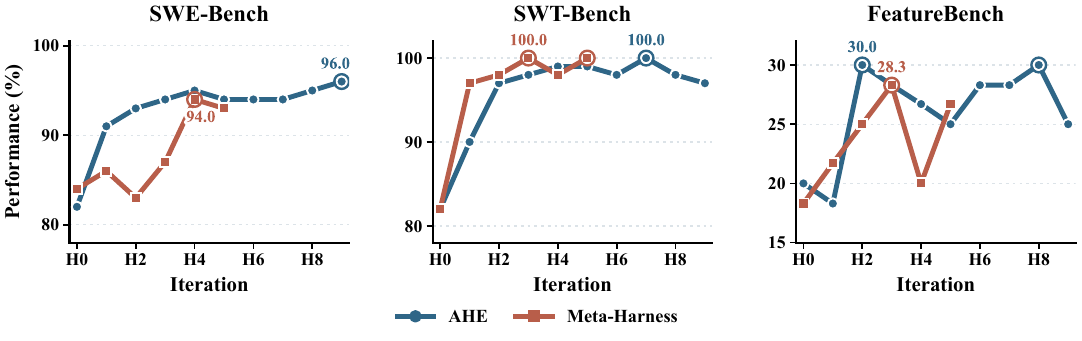}
  \caption{Harness evolution on single dataset. Performance of AHE and Meta-Harness over successive evolution rounds on SWE-Bench, SWT-Bench, and FeatureBench. The annotations report the best-performing harness identified during each evolution process.}
  \label{fig:single_dataset}
\end{figure*}

\begin{figure*}[t]
  \centering
  \includegraphics[width=\linewidth]{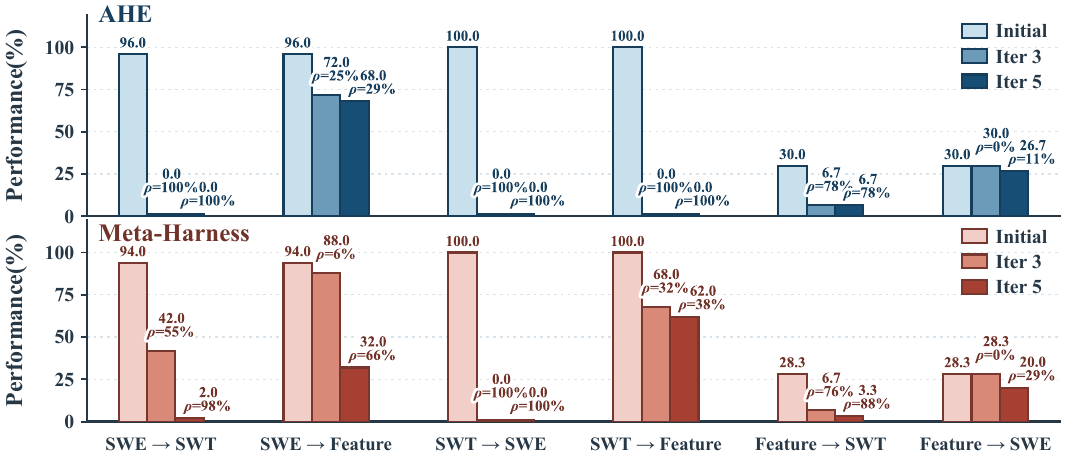}
  \caption{Catastrophic forgetting during continual harness self-evolution across six directions. For each direction, the harness evolved on a single dataset is continuously evolved on another dataset, with its earlier-dataset performance evaluated initially and after 3 and 5 additional continual evolution iterations (Initial, Iter 3, and Iter 5). Labels on evolved bars show the performance score and relative performance drop ($\rho$).}
  \label{fig:cross_dataset}
\end{figure*}

\subsection{RQ1: Catastrophic Forgetting During Continual Harness Self-Evolution}

To investigate whether continual harness self-evolution leads to catastrophic forgetting, we examine how it affects performance on previously encountered tasks in this RQ.

\textbf{Setting.} In our experiments, we set up continual harness self-evolution on two datasets of different types. The conclusions obtained from this experimental setting can also be generalized to multiple sequential datasets: an evolution method that causes catastrophic forgetting on adjacent tasks will likewise cause forgetting in continual evolution across multiple datasets. We first perform self-evolution independently on SWE-Bench, SWT-Bench, and FeatureBench. After that, we select the best-performing harness obtained during each evolution process as the initial harness, transfer it to another dataset for further evolution, and evaluate its performance degradation on the previous dataset. When multiple harnesses achieve the same highest score, we select the first harness that reaches this score. For example, for AHE on FeatureBench, we select the harness obtained after the third round as an initial for continual evolution, which achieves a score of 30.0\%.

\textbf{Results.} The experimental results derive the following key findings:

\textbf{\ding{182} Both self-evolution methods demonstrate substantial performance improvements when evolving on a single dataset.} As shown in Figure~\ref{fig:single_dataset}, $H_0$ denotes the initial harness of Mini-SWE-Agent, while $H_i$ denotes the harness after the $i$-th round of evolution. For example, AHE improves the performance on SWE-Bench from 82.0\% initially to 96.0\% after nine rounds of evolution, while Meta-Harness achieves a final score of 94.0\%. On SWT-Bench, both methods improve the performance from the initial score of 82.0\% to 100.0\% on the randomly sampled 50 tasks. On FeatureBench, both methods eventually achieve an improvement of more than 10 percentage points. We circle the highest-scoring point in each line plot in Figure~\ref{fig:single_dataset}, highlighting the best-performing harness.

\textbf{\ding{183} Existing self-evolution methods struggle to preserve previously acquired capabilities during continual harness self-evolution.} As shown in Figure~\ref{fig:cross_dataset}, almost every continual harness self-evolution direction results in substantial performance degradation on previously encountered tasks. For example, when evolving from SWE-Bench to SWT-Bench, AHE and Meta-Harness suffer performance degradation of 100\% and 98\% on SWE-Bench, respectively, after five rounds of evolution. In other words, after evolving on SWT-Bench, the resulting agents are almost completely unable to solve the SWE-Bench tasks. Similarly, in the SWE $\rightarrow$ Feature setting, the performance of AHE and Meta-Harness decreases by 29\% and 66\%, respectively. 

\textbf{\ding{184} The extent of catastrophic forgetting varies substantially across datasets and evolution directions.} In the software engineering domain, SWT-Bench~\citep{DBLP:conf/nips/MundlerMHV24} evaluates agents on test case generation, whereas SWE-Bench~\citep{DBLP:conf/iclr/JimenezYWYPPN24} and FeatureBench~\citep{DBLP:conf/iclr/ZhouZWHWHYWPFTZ26} require agents to modify the source code of existing repositories. Therefore, SWT-Bench is relatively less related to the other two datasets, while SWE-Bench and FeatureBench involve more closely related types of tasks. As shown in Figure~\ref{fig:cross_dataset}, evolution between SWE-Bench and FeatureBench causes relatively smaller performance degradation on the previously encountered dataset, whereas evolution between SWT-Bench and the other two datasets leads to substantially larger degradation. These results suggest that task relatedness may play an important role in mitigating catastrophic forgetting during continual harness self-evolution. Evolving on a substantially different task type can introduce modifications that are highly detrimental to capabilities acquired from the previous tasks, resulting in more severe forgetting.

\begin{rpsinsight}{Answer to RQ1}
Existing harness self-evolution methods can improve performance on an individual task type, but often suffer catastrophic forgetting in the continual evolution. The extent of forgetting is closely associated with the dataset and the direction of evolution.
\end{rpsinsight}

\subsection{RQ2: Harness Components Associated with Catastrophic Forgetting}

\begin{figure*}[t]
    \centering
    \includegraphics[width=\linewidth]{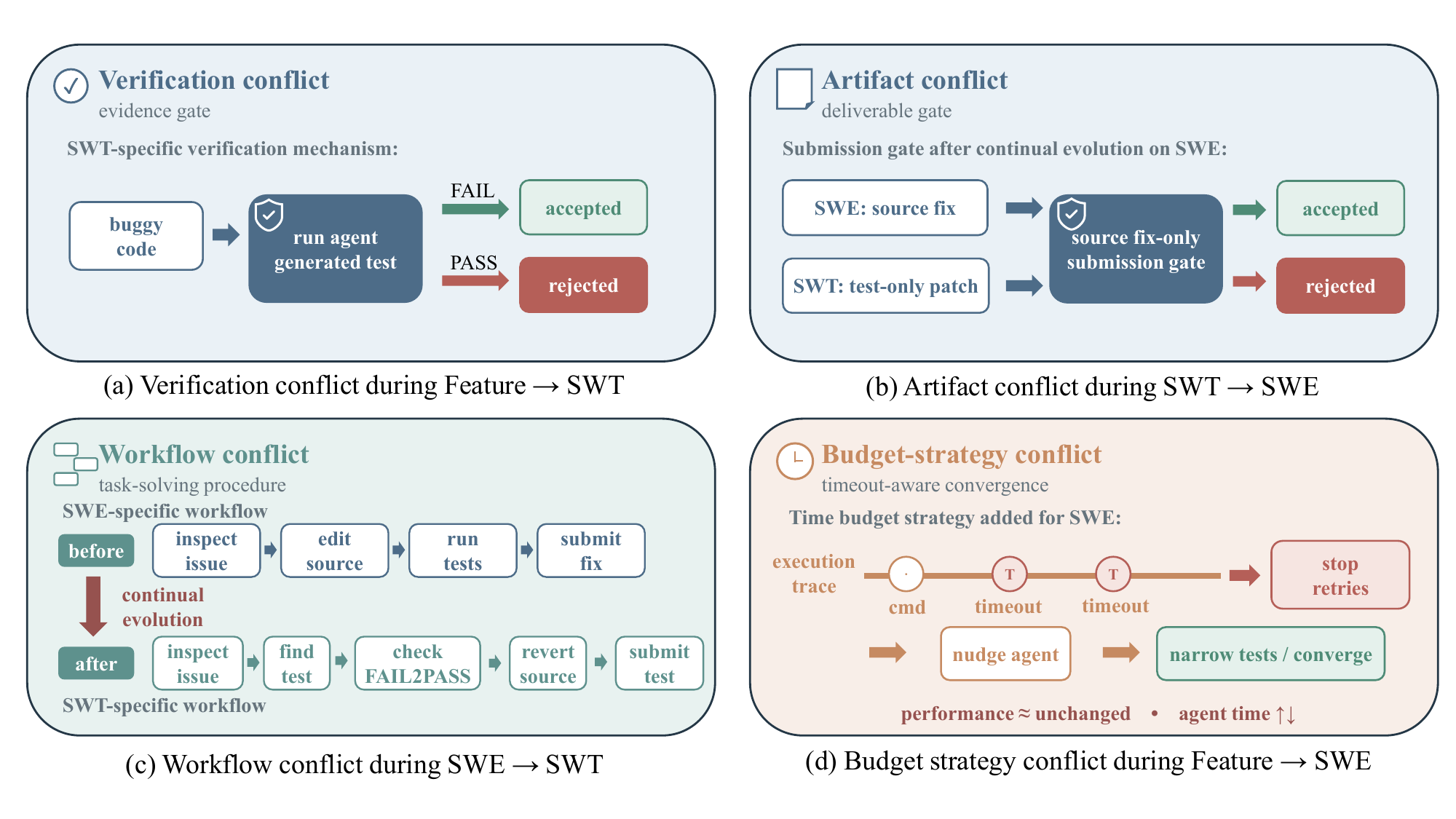}
    \caption{Examples of Harness Conflicts in Continual Harness Self-Evolution.}
    \label{fig:harness_conflict}
\end{figure*}

\subsubsection{Categorization of Harness Conflicts}

In this RQ, we further analyze the reasons why harness modifications lead to performance degradation on previously encountered tasks.

\textbf{Setting.} To characterize the types of harness conflicts introduced during continual harness self-evolution, we conduct a two-stage analysis for each evolution direction in Figure~\ref{fig:cross_dataset}. First, we extract the code diff between the initial harness and the harness obtained after five rounds of continual evolution, and manually examine the modifications to identify the types of conflicts introduced by the evolution process. We then analyze a total of 520 execution trajectories and identify the tasks that exhibit a \textit{pass-to-fail} transition after continual harness self-evolution. For these cases, we analyze their execution trajectories in detail and determine which types of harness conflicts are associated with the observed failures. Finally, we calculate the proportion of each conflict type among these \textit{pass-to-fail} cases to quantify their prevalence in the observed performance degradation.

\textbf{Results.} We identify three major types of modifications that can introduce such conflicts: \textit{verification conflicts},  \textit{workflow conflicts} and \textit{budget strategy}. Among the pass-to-fail cases, verification conflicts account for 87.1\%, workflow conflicts account for 8.2\%, and budget conflicts account for 0.6\%. The remaining cases are not associated with a specific conflict type.

\textbf{\ding{182} \textit{Verification conflicts} arise when the harness introduces checks, constraints, or reminders that are tailored to the requirements of a particular dataset.} The harness may determine what evidence should be inspected, what conditions must be satisfied before submission, or what artifact should be delivered. These interventions directly influence the agent's execution trajectory and can conflict with the requirements of previously encountered tasks. Figure~\ref{fig:harness_conflict} (a) shows a normal example during the continual evolution from FeatureBench to SWT-Bench. It means that if a newly generated test already passes on the buggy code, it does not provide fail-to-pass evidence, and the submission is therefore blocked. This specific modification of SWT-Bench can negatively affect performance on previous tasks.

Among the verification conflicts, \textit{Artifact conflicts} are the most prominent. When a harness evolves from one type of task to another, artifact-specific constraints may be incorporated into the harness, such as enforcing the expected artifact type at submission time. These modifications can conflict with the requirements of previous tasks, causing the agent to produce an inappropriate artifact and leading to task failures. As shown in Figure~\ref{fig:harness_conflict} (b), in the evolution from SWT-Bench to SWE-Bench, the required submission artifact switches from a test-only patch to a source-code fix, accompanied by a reversal in the polarity of the submission gate.

\textbf{\ding{183} \textit{Workflow conflicts} arise from differences in the procedures required to solve different types of tasks.} Such differences are typically reflected in the prompts or workflow-related helpers of the harness. As shown in Figure~\ref{fig:harness_conflict} (c), during the evolution from SWE-Bench to SWT-Bench, the prompt is modified to change the agent's problem-solving workflow. Instead of directly modifying the source code after locating the issue, the agent identifies the relevant test file, writes a regression test, verifies that it fails before the fix and passes after the fix, and finally reverts the temporary source-code modifications. While these modifications may benefit the current task type, they can interfere with workflows that are more appropriate for previously encountered tasks.

\textbf{\ding{184} \textit{Budget strategy} concerns how the harness manages the limited time and computational resources available to the agent.} Such strategies include reminders based on elapsed wall-clock time or the number of steps, as well as guards for repeated command failures, timeouts, or slow execution. During continual harness self-evolution, these budget-related mechanisms may be modified to accommodate the resource requirements of the current tasks. Figure~\ref{fig:harness_conflict} (d) shows a representative modification to the budget strategy introduced during the evolution from FeatureBench to SWE-Bench. After repeated command timeouts, the harness prompts the agent to stop repeatedly retrying the same commands, narrow the scope of testing, and converge toward a solution more quickly.

\subsubsection{Impact and Case Study of Different Harness Conflict Types}

\input{tables/1_ablation}

The prevalence of different conflict types does not necessarily reflect their impact on agent performance. To examine their individual effects, we introduce representative modifications associated with each conflict type into the harness and evaluate the resulting performance against that of the original harness. We further analyze failure cases to understand how each type of conflict leads to task failures.

\textbf{Setting.} We take the AHE evolution from SWE-Bench to FeatureBench as an example. We add the workflow and time budget-related modifications identified during continual harness self-evolution to the best performing harness previously evolved on SWE-Bench, and examine the resulting performance impact of each modification on SWE-Bench. Since SWE-Bench and FeatureBench do not differ substantially in terms of their submission artifacts, we instead use the evolution from SWE-Bench to SWT-Bench as the source of modifications for verification conflicts.

\textbf{Results.} As shown in Table~\ref{tab:conflict_type_ablation}, we find that introducing a mandatory artifact constraint immediately reduces the performance on SWE-Bench to 0.0. During continual harness self-evolution from SWE-Bench to SWT-Bench, the different deliverable requirements reverse the submission logic: the evolved harness automatically checks the agent's modifications before submission and rejects any solution that modifies files beyond the test files. This constraint is suitable for SWT-Bench but incompatible with SWE-Bench, where source-code modifications are required to fix the issue. 

Workflow conflicts also have a substantial impact, reducing the performance from 96.0 to approximately 66.0. During continuous evolution from SWE-Bench to FeatureBench, the evolver substantially modified the agent's workflow in the system prompt, while also introducing hard constraints at the harness level to enforce this workflow. In particular, the evolved harness intercepts network-related commands such as pip, curl, and git, preventing the agent from accessing external resources. As a result, the agent loses access to potentially useful external resources, leading to a substantial degradation in SWE-Bench performance. 
Moreover, following the FeatureBench-specific workflow leads the agent to perform a large number of unnecessary exploratory operations, increasing the average task execution time by 235\%.

In contrast, budget strategies have little impact on task performance, with their effects mainly reflected in changes in task execution efficiency. As shown in Table~\ref{tab:conflict_type_ablation}, introducing the budget strategy learned from FeatureBench increases the agent's execution time by 79\%.

\begin{rpsinsight}{Answer to RQ2}
The conflicts in the harness can be categorized into three types: verification conflicts, workflow conflicts and budget strategy. Their impacts vary substantially, with artifact and workflow conflicts causing the most severe performance degradation, and the budget strategy primarily reflecting the agent's execution time.
\end{rpsinsight}

%% file: tables/1_ablation.tex
\begin{table}[H]
  \centering
  \caption{Impact of Individual Harness Conflict Types on
  SWE-Bench Performance and Agent Execution Time.}
  \label{tab:conflict_type_ablation}
  \small
  \renewcommand{\arraystretch}{1.15}
  \setlength{\tabcolsep}{6pt}
  \begin{tabular}{lrr}
    \toprule
    \textbf{Harness Configuration}
    & \textbf{SWE Performance (\%)}
    & \textbf{Agent Time (s)} \\
    \midrule
    Original Best Harness & 96.0 & 245.5 \\
    + Verification Conflict & 0.0 ($\downarrow$100\%) & 415.7 ($\uparrow$69\%) \\
    + Workflow Conflict & 66.0 ($\downarrow$31\%) & 821.5 ($\uparrow$235\%) \\
    + Budget Strategy & 96.0 ($\downarrow$0\%) & 439.7 ($\uparrow$79\%) \\
    \bottomrule
  \end{tabular}
\end{table}

%% file: sections/5_related_work.tex
\section{Related Work}
\label{sec:related}
\subsection{Harness Self-Evolution}

Recent work on harness self-evolution focuses on improving the runtime system surrounding the foundation model~\citep{DBLP:journals/corr/abs-2604-25850, DBLP:journals/corr/abs-2603-28052, DBLP:journals/corr/abs-2606-09498, DBLP:journals/corr/abs-2606-14249}. Given a set of tasks, an agent first produces a collection of execution traces. The evolution process then analyzes these traces to identify successful behaviors or diagnose the root causes of failures, and subsequently makes attributable modifications to the agent's harness based on the accumulated evidence. A substantial body of work focuses on evolving individual harness components, including prompts~\citep{DBLP:conf/iclr/AgrawalTSZKOSSR26, DBLP:conf/iclr/ZhangHUMHKRWJLT26}, memory~\citep{DBLP:journals/corr/abs-2508-16153, DBLP:conf/ecai/ChhikaraKASY25}, and skills~\citep{DBLP:journals/corr/abs-2604-01687, DBLP:journals/corr/abs-2602-08234}, while other approaches optimize the harness as a whole, such as AHE~\citep{DBLP:journals/corr/abs-2604-25850}, Meta-Harness~\citep{DBLP:journals/corr/abs-2603-28052}, Self-Harness~\citep{DBLP:journals/corr/abs-2606-09498} and HarnessX~\citep{DBLP:journals/corr/abs-2606-14249}. 

However, existing studies on harness self-evolution primarily evaluate whether the evolution process improves agent capabilities on a single type of task. This line of work does not explicitly consider whether existing methods can continuously evolve harnesses across multiple kinds of tasks.

\subsection{Continual Learning for Agents}
In recent years, research on continual learning has gradually expanded from continual learning for LLMs to continual learning for LLM-based agents~\citep{DBLP:journals/pami/ZhengSCLZLYM26}. For example, Voyager~\citep{DBLP:journals/tmlr/WangX0MXZFA24} continuously accumulates and reuses skills during interaction, while CER~\citep{DBLP:conf/acl/LiuSNY25} and LifeMem~\citep{DBLP:journals/corr/abs-2609-12655} focus on retaining and leveraging historical experiences to assist agents in completing current tasks. Several benchmarks have also been proposed to evaluate continual learning for agents. LifelongAgentBench~\citep{DBLP:journals/corr/abs-2505-11942} and EvoAgentBench~\citep{DBLP:journals/corr/abs-2607-05202} focus on skill-based agents, while AGENTCL~\citep{DBLP:journals/corr/abs-2606-02461}, Continual Learning Bench~\citep{DBLP:journals/corr/abs-2606-05661}, and SWE-Bench-CL~\citep{DBLP:journals/corr/abs-2507-00014} consider task ordering and experience accumulation over time, primarily evaluating memory-augmented agents.

However, relatively little attention has been paid to the continual evolution of the agent harness as a whole, where multiple components can be jointly modified based on accumulated execution experience. 

%% file: sections/6_conclusion.tex
\section{Conclusion}
\label{sec:conclu}
In this paper, we study continual harness self-evolution and identify catastrophic forgetting as a central challenge. We evaluate two representative harness self-evolution methods across three software engineering benchmarks and six cross-task evolution directions. While both methods improve performance on individual tasks, their performance often degrades substantially when evolution spans different tasks. Further analysis identifies three types of harness-level conflicts that contribute to catastrophic forgetting: verification, workflow, and budget-strategy conflicts. These findings highlight the challenges of evolving coding agents reliably across diverse software engineering tasks.

%% file: appendix/1_implementation_details.tex
\section{Implementation Details}
\label{app:details}

\input{tables/2_settings}

%% file: tables/2_settings.tex
\begin{table}[H]
  \centering
  \caption{Configuration for reproducibility.}
  \label{tab:settings}
  \small
  \renewcommand{\arraystretch}{1.15}
  \setlength{\tabcolsep}{5pt}
  \begin{tabularx}{\linewidth}{lX}
    \toprule
    \textbf{Item} & \textbf{Setting to report} \\
    \midrule
    Initial harness & Mini-SWE-Agent v2.4.6, extended with a tool registry \\
    Evolvable surface & src/minisweagent/config/mini.yaml (prompts), src/minisweagent/agents/** (control flow), src/minisweagent/models/utils/** (tools, parsers, observations) \\
    Model & DeepSeek-V4-Flash-0731 \\
    Rollouts & 2 per task per round \\
    Sandbox & E2B, 1 h sandbox lifetime \\
    Evolution rounds & AHE: 9 rounds of evolution (10 evaluated harnesses); Meta-Harness: 5 rounds \\
    Datasets & SWE-bench Verified (50), SWT-bench paraphrased (50), FeatureBench (30) \\
    \bottomrule
  \end{tabularx}
\end{table}

%% file: appendix/2_discussions.tex
\section{Discussion and Case Study for Dataset Settings}
\label{app:disc}

In Section~\ref{sec:exp_eval}, we described a specific setting adopted in our experiments: we paraphrased the fixed linguistic pattern shared by the instances in SWT-Bench while preserving their original semantics. In this section, we further discuss and validate the rationale behind this experimental setting.

\begin{figure}[t]
  \centering

  \begin{minipage}{\linewidth}
    \centering
    \includegraphics[width=\linewidth]{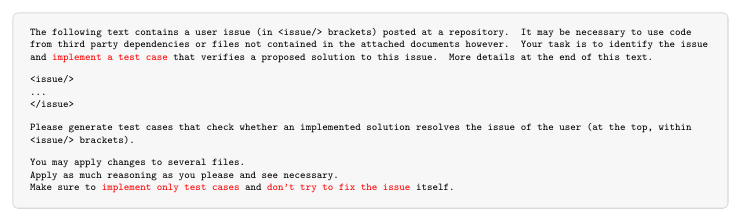}
    
    \smallskip
    \textbf{(a)} SWT-Bench prompt template before paraphrasing.
  \end{minipage}

  \vspace{6pt}

  \begin{minipage}{\linewidth}
    \centering
    \includegraphics[width=\linewidth]{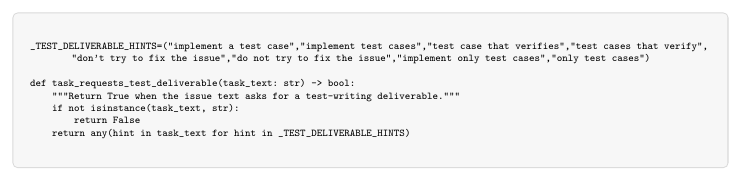}
    
    \smallskip
    \textbf{(b)} Test-deliverable hints in the evolved harness.
  \end{minipage}

  \vspace{6pt}

  \begin{minipage}{\linewidth}
    \centering
    \includegraphics[width=\linewidth]{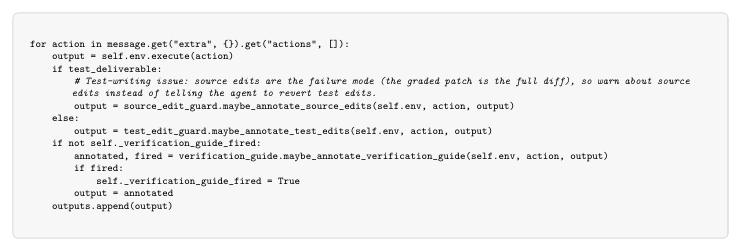}
    
    \smallskip
    \textbf{(c)} Submission-condition branch introduced during evolution from SWE-Bench to SWT-Bench.
  \end{minipage}

  \caption{Case study of the harness modification introduced during evolution from SWE-Bench to SWT-Bench.}
  \label{fig:discussion_swt}
\end{figure}

If the original SWT-Bench prompts are kept unchanged, as shown in Figure~\ref{fig:discussion_swt}(a), we observe that the Meta-Harness evolution from SWE-Bench to SWT-Bench introduces a conditional branch for determining the submission artifact, as shown in Figure~\ref{fig:discussion_swt}(c). Specifically, if the harness identifies the current task as a code modification task, it requires the agent to submit source code; otherwise, it requires the agent to submit a test file. This behavior is a positive signal for mitigating catastrophic forgetting, as it allows the evolved harness to distinguish between different types of tasks and preserve the corresponding submission requirements.

However, the condition used to determine \texttt{test\_deliverable} is constructed entirely based on the fixed linguistic template used in SWT-Bench, as shown in Figure~\ref{fig:discussion_swt}(b). Such an evolution pattern is clearly unfavorable for generalization, as the learned condition is closely tied to the surface form of the SWT-Bench prompts rather than the underlying task semantics.

After paraphrasing the SWT-Bench prompts, we no longer observe similar template-dependent conditional branches during self-evolution. This observation suggests that prompt paraphrasing can mitigate the risk of self-evolution exploiting dataset-specific linguistic patterns, encouraging the evolved harness to rely more on task-level characteristics.

